\documentclass[useAMS,usenatbib,usegraphicx]{mn2e}
\usepackage[dvips]{graphics}
\usepackage{colordvi}
\usepackage{rotating}
\usepackage{ifthen}
\usepackage{amssymb}
\usepackage{epsfig}
\usepackage{wrapfig}
\usepackage{longtable}
\usepackage{times}

\title[On the binarity of Be stars]{The binary fraction and mass-ratio of Be and B stars: a comparative VLT/NACO study\thanks{Based on observations collected at the European Southern Observatory
(ESO), Paranal, Chile under programme ID 075.C-0475(A).}
}

\author[Ren\'e  Oudmaijer and Andy Parr ] 
{Ren\'e D. Oudmaijer and Andrew M. Parr \\ 
School of Physics and Astronomy, EC Stoner Building, University
of Leeds, Leeds LS2 9JT, UK\\
}

\begin{document}

\date{Accepted 2010 March 2. Received 2010 February 25; in original form 2010 January 10}

\pagerange{\pageref{firstpage}--\pageref{lastpage}} \pubyear{2004}

\maketitle

\label{firstpage}

\begin{abstract} 

In order to understand the formation mechanism of the disks around Be
stars it is imperative to have a good overview of both the differences
and similarities between normal B stars and the Be stars. In this
paper we address binarity of the stars. In particular, we investigate
a previous report that there may be a large population of
sub-arcsecond companions to Be stars.  These had hitherto not been
detected due to a combination of a limited dynamic range and spatial
resolution in previous studies.  We present the first systematic,
comparative imaging study of the binary properties of matched samples
of B and Be stars observed using the same equipment. We obtained high
angular resolution (0.07-0.1 arcsec) {\it K} band Adaptive Optics data
of 40 B stars and 39 Be stars. The separations that can be probed
range from 0.1 to 8 arcsec (corresponding to 20-1000 AU), and
magnitude differences up to 10 magnitudes can in principle be covered.
After excluding a few visual binaries that are located in regions of
high stellar density, we detect 11 binaries out of 37 Be targets
(corresponding to a binary fraction of 30 $\pm$ 8 \%) and 10 binaries
out of 36 B targets (29 $\pm$ 8\%).  Further tests demonstrate that
the B and Be binary systems are not only similar in frequency but also
remarkably similar in terms of binary separations, flux differences
and mass ratios.  The minimum physical separations probed in this
study are of order 20 AU, which, combined with the similar binary
fractions, indicates that any hypotheses invoking binary companions as
responsible for the formation of a disk need the companions to be
closer than 20 AU.  Close companions are known to affect the
circumstellar disks of Be stars, but as not all Be stars have been
found to be close binaries, the data suggest that binarity can not be
responsible for the Be phenomenon in all Be stars.  Finally, the
similarities of the binary parameters themselves also shed light on
the Be formation mechanism. They strongly suggest that the initial
conditions that gave rise to B and Be star formation must, to all
intents and purposes, be similar. This in turn indicates that the Be
phenomenon is not the result of a different star formation mechanism.

\end{abstract}

\begin{keywords} 
  techniques: high angular resolution - stars: binaries - stars: emission line, Be - stars: statistics
\end{keywords}

\begin{table*}
\begin{center}
\begin{tabular}{|lcclrl|llrlrl|}                                
\hline
\hline
\multicolumn{6}{c}{the B stars} &  \multicolumn{6}{c}{the Be stars}  \\
\hline
HR	&        RA	        & Dec &{\it V}	&{\it B$-$V} &	SpType &HR	&        RA	        & Dec &{\it V}	&{\it B$-$V} &	SpType \\
\hline
1619	&	05 02 00.0	&+01 36 32	& 6.24	&-0.04	&  B8V    &  1508	&	04 44 05.3	&-08 30 13	& 5.90	&-0.11	&  B2Ve    \\
1753	&	05 19 17.5	&-18 31 12	& 6.36	&-0.16	&  B3V    &  1956	&	05 39 38.9	&-34 04 27	& 2.64	&-0.12	&  B7IVe    \\
1855	&	05 31 55.8	&-07 18 05	& 4.62	&-0.26	&  B0V    &  2142	&	06 04 13.5	&-06 42 33	& 5.21	&-0.06	&  B2Ven    \\
1892	&	05 35 23.2	&-04 50 18	& 4.59	&-0.19	&  B1V    &  2170	&	06 07 03.7	&-34 18 43	& 5.83	&-0.13	&  B4IVe    \\
2056	&	05 53 06.9	&-33 48 05	& 4.87	&-0.15	&  B5V    &  2231	&	06 15 40.1	&+06 03 58	& 6.07	&-0.13	&  B6Ve    \\
2266	&	06 18 13.7	&-19 58 01	& 5.52	&-0.18	&  B2V    &  2288	&	06 20 36.3	&-34 08 38	& 5.53	&-0.19	&  B1.5Ve    \\
2344	&	06 27 57.6	&-04 45 44	& 5.06	&-0.17	&  B2V    &  2309	&	06 24 20.5	&-12 57 45	& 6.12	&-0.08	&  B5Ve    \\
2544	&	06 50 36.9	&-25 46 41	& 6.33	&-0.18	&  B2V    &  2356	&	06 28 49.0	&-07 01 58	& 4.60	&-0.10	&  B3Ve    \\
2704	&	07 09 42.9	&-25 13 52	& 5.69	&-0.16	&  B2.5IV    &  2357	&	06 28 49.5	&-07 02 04	& 5.40	&-0.07	&  B3ne    \\
3192	&	08 09 01.6	&-19 14 42	& 4.40	&-0.15	&  B5IV    &  2358	&	06 28 49.5	&-07 02 04	& 5.60	&     	&  B3e    \\
3240	&	08 13 58.4	&-36 19 21	& 5.08	&-0.19	&  B1.5IV    &  2397	&	06 32 39.0	&-32 01 50	& 5.69	&-0.17	&  B2IVe    \\
3440	&	08 39 23.8	&-53 26 23	& 5.48	&-0.16	&  B5V    &  2412	&	06 31 58.4	&-58 45 15	& 5.70	&-0.06	&  B9Ve    \\
3665	&	09 14 21.9	&+02 18 51	& 3.88	&-0.06	&  B9.5V    &  2577	&	06 54 42.1	&-01 45 23	& 6.21	& 0.56	&  B3IVe+K2II    \\
3734	&	09 22 06.8	&-55 00 39	& 2.50	&-0.18	&  B2IV-V    &  2690	&	07 07 22.6	&-23 50 25	& 5.71	&-0.10	&  B2IVe    \\
3990	&	10 08 56.3	&-51 48 40	& 4.86	&-0.12	&  B3IV    &  2749	&	07 14 48.7	&-26 46 22	& 3.85	&-0.17	&  B2IV-Ve    \\
4292	&	10 59 59.4	&-43 48 26	& 5.81	&-0.08	&  B8-9V    &  2825	&	07 24 40.1	&-16 12 04	& 5.33	&-0.05	&  B2.5IVe    \\
4403	&	11 24 22.1	&-42 40 09	& 6.12	&-0.18	&  B2IV-V    &  2968	&	07 39 58.0	&-37 34 46	& 6.00	&-0.04	&  B6IVe    \\
4590	&	12 00 51.2	&-19 39 32	& 5.26	&-0.20	&  B1.5V    &  3213	&	08 09 43.2	&-47 56 15	& 5.23	&-0.21	&  B1IVe    \\
4648	&	12 13 25.3	&-38 55 45	& 5.76	&-0.13	&  B4IV    &  3330	&	08 25 31.0	&-51 43 41	& 5.17	&-0.16	&  B2Ve    \\
4940	&	13 06 16.7	&-48 27 49	& 4.71	&-0.14	&  B5V    &  3488	&	08 46 23.8	&-41 07 32	& 6.21	&-0.05	&  B9Ve    \\
5231	&	13 55 32.4	&-47 17 18	& 2.55	&-0.22	&  B2.5IV    &  3670	&	09 13 34.5	&-47 20 19	& 5.92	&-0.05	&  B9Ve    \\
5395	&	14 26 08.2	&-45 13 17	& 4.56	&-0.15	&  B2IV    &  3745	&	09 26 44.8	&-28 47 15	& 6.10	&-0.10	&  B6Ve    \\
5559	&	14 56 32.0	&-47 52 45	& 5.64	&-0.04	&  B9V    &  3858	&	09 41 17.0	&-23 35 30	& 4.77	&-0.12	&  B6Ve    \\
5614	&	15 05 47.7	&-25 47 23	& 6.67	&-0.01	&  B8V    &  3946	&	09 59 06.1	&-23 57 01	& 6.21	&-0.10	&  B4Ve    \\
5708	&	15 22 40.9	&-44 41 22	& 3.37	&-0.18	&  B2IV-V    &  4018	&	10 13 01.3	&-59 55 05	& 6.10	&-0.08	&  B4Ve    \\
5805	&	15 38 32.7	&-39 09 39	& 6.57	&-0.07	&  B9V    &  4460	&	11 34 45.7	&-54 15 51	& 4.62	&-0.08	&  B9Ve    \\
5934	&	15 57 40.4	&-20 58 59	& 5.85	& 0.02	&  B3V    &  4823	&	12 41 56.6	&-59 41 09	& 4.93	&-0.04	&  B6IVe    \\
5993	&	16 06 48.4	&-20 40 09	& 3.96	&-0.04	&  B1V    &  5193	&	13 49 37.0	&-42 28 26	& 3.04	&-0.17	&  B2IV-Ve    \\
6080	&	16 20 32.6	&-39 25 51	& 6.12	&-0.07	&  B9V    &  5311	&	14 13 39.9	&-54 37 33	& 6.11	& 0.05	&  B5Ve    \\
6174	&	16 38 26.2	&-43 23 55	& 5.83	&-0.02	&  B2.5IV    &  5661	&	15 16 36.4	&-60 54 14	& 5.73	&-0.08	&  B0.5Ve    \\
6316	&	17 01 52.7	&-32 08 37	& 5.03	&-0.10	&  B8V    &  5683	&	15 18 32.0	&-47 52 30	& 4.27	&-0.08	&  B8Ve    \\
6494	&	17 27 37.5	&-29 43 28	& 6.00	& 0.00	&  B9IV    &  6519	&	17 31 25.0	&-23 57 46	& 4.81	& 0.00	&  B9.5Ve    \\
6628	&	17 49 10.5	&-31 42 12	& 4.83	&-0.04	&  B8V    &  6621	&	17 48 27.8	&-26 58 30	& 6.35	& 0.12	&  B4IVe    \\
6692	&	17 58 39.1	&-28 45 33	& 6.01	&-0.08	&  B3IV    &  6712	&	18 00 15.8	&+04 22 07	& 4.64	&-0.03	&  B2Ve    \\
6893	&	18 25 54.6	&-33 56 43	& 6.30	&-0.08	&  B5IV    &  6873	&	18 21 28.5	&+05 26 09	& 6.13	&-0.04	&  B3Ve    \\
6960	&	18 33 57.8	&-33 01 00	& 5.28	&-0.11	&  B2IV-V    &  6881	&	18 23 12.2	&-12 00 53	& 5.73	& 0.01	&  B8IV-Ve    \\
7169	&	19 01 03.2	&-37 03 39	& 6.69	&     	&  B9V    &  7249	&	19 08 16.7	&-19 17 24	& 5.54	&-0.11	&  B2Ve    \\
7170	&	19 01 04.3	&-37 03 43	& 6.40	&-0.03	&  B8V-IV    &  7554	&	19 50 17.5	&+07 54 09	& 6.51	&-0.10	&  B2.5IVe    \\
7348	&	19 23 53.2	&-40 36 58	& 3.97	&-0.10	&  B8V    &  8402	&	22 03 18.9	&-02 09 19	& 4.69	&-0.06	&  B7IVe    \\
7709	&	20 11 10.1	&-08 50 32	& 6.49	&-0.15	&  B1V    &  8539	&	22 25 16.6	&+01 22 39	& 4.66	&-0.03	&  B1Ve    \\
8141	&	21 18 11.1	&-04 31 10	& 5.82	&-0.13	&  B5V    & \\
\hline
 \hline
\end{tabular}
\caption{The target sample.  Unless otherwise noted in the text,
photometry and spectral types are taken from the Bright Star Catalogue
\citep{bsc}, the coordinates are J2000.  HR 2577 is an unresolved spectroscopic binary.\label{logtarget}
}
\end{center}                                                  
\end{table*}

\section{Introduction}

Be stars are rapidly rotating stars surrounded by an ionized
circumstellar disk, whose emission is best known for the distinct
doubly peaked H$\alpha$ lines. About 15\% of all B stars belong to the
Be category \citep[chap. 9]{jaschek_book}, making them a significant
component of the hot, massive star population.  However, the origin of
the circumstellar disks around Be stars is still an unsolved mystery
(see e.g. the review by \citealt{porter_review}).

Binarity has sometimes been proposed to be responsible for the Be
phenomenon. For example, \citet{harmanec_2002} investigate the effects of a
detached, secondary component's gravitational field on the mass ejection into the
equatorial plane from the primary.  \citet{gies_1998} discuss the rotational
history of $\phi$ Per, and find the (compact) companion has had a significant
tidal effect on the central Be star. In addition, it should be noted that
companions are also known to affect the disk's properties such as its density
law or even size (e.g. \citealt{jones_2008} on $\kappa$ Dra).

Most models based on binarity have in common that close binaries with
separations of order AU, are responsible for the formation of a disk (for an
overview, see \citealt{Neg_2007}). When one considers the more general case of
binaries, it is plausible that disks are affected during the star formation
stages by the presence of a companion star, even in the case of a wide binary.
However, to be able to address binarity as a crucial ingredient for the
existence of Be stars, we need to know whether Be stars are found in binary
systems more often than normal B stars.  If so, this would indicate that
binarity must have something to do with the production of the conditions
required for disk formation -- either directly, where the tidal forces acting
within a (close) binary system result in a disk around a mass losing object,
or indirectly, where the influence of the binary companion allows the existing
disk to survive.

\begin{figure}
\includegraphics[width=0.4\textwidth,angle=-90]{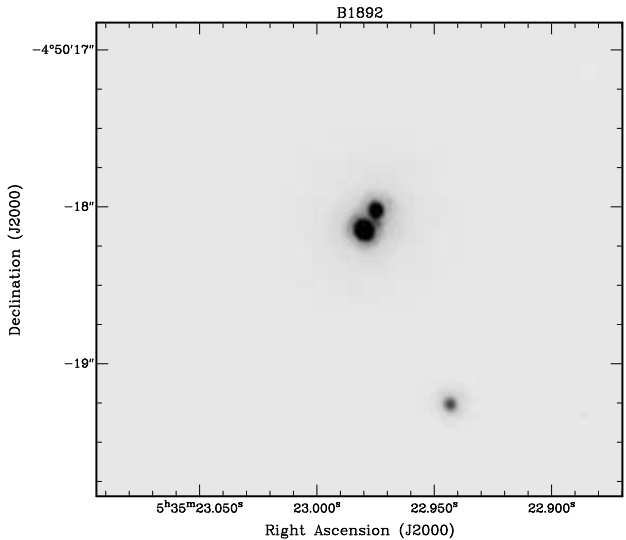}
\caption{Example of the data. Shown in this 3$\times$3 arcsec image is the
  known B-type binary HR 1892. The previously known and catalogued companion is
  located at a distance of 1.6 arcsec towards the South-West. The close
  companion at 0.15 arcsec (double the spatial resolution in this image)
  in the North-West direction is a new detection. \label{h1892}}
\end{figure}

The most straightforward manner to determine whether the Be binary fraction is
significantly different from that of ``normal'' B-type stars and thereby to
assess whether existing binary models are viable in the first place is to
establish whether Be stars are more frequently found in binaries than normal B
stars.  The most recent comparative study on Be binarity was performed 25
years ago by \citet{abt_1984}. In this study, incorporating the spectroscopic
survey of Be and B stars by \citealt{abt_1978}, and including both spectroscopic
and visual binaries, they found the Be binary fraction to be 30\%, similar to
that of B stars. However, this paper was essentially a compilation of the
then, inevitably incomplete, literature on visual binaries and was never
repeated by a homogeneous dedicated investigation. Later studies, such as the
Speckle interferometric study by \citet{mason_1997} concentrated on Be stars
alone. Despite their high angular resolution, Mason et al. did not discover
many new binary systems. This is most likely due to their limited dynamic
range -- only companions which were fainter by up to 3 magnitudes than the
primary could be detected.

The original motivation for the current study was the unexpected result that
in an investigation into the spectro-astrometric signatures of the H$\alpha$
emission lines of a random sample of 8 Be stars, 5 were found to be a binary
with separations at the sub-arcsecond to arcsecond scale
(\citealt{baines_thesis}, see also \citealt{oudmaijer_2008}).   The parameter
space probed by the spectro-astrometry (from 0.1 arcsec to 1-2 arcsec, up to 6
magnitudes difference, \citealt{baines_2006}) was not covered in any previous
study into the binarity of Be stars (see references above), and
observationally this was a completely unconstrained problem.  If true, the
large fraction of (sub-)arcsecond Be star binaries, would have important
implications for our understanding of Be stars. Obviously, if wide
binarity could be linked to the Be star phenomenon it would require new
physics or new models as the effect from distant companions will be subtle.
Alternatively, low number statistics could have mimicked a high binary
fraction, whereas, in reality, this may not be the case.  A follow-on
study with more objects matched by a sample of normal B stars should allow us
to decide on the issue. Whereas the technique of spectro-astrometry is very
powerful in detecting binaries amongst {\it emission}-line stars, it is not as
sensitive when considering normal, absorption line, B-type stars
\citep{baines_2006}. In order to do a comparative study, we need to employ a
technique that is both powerful and applicable to both types of star.

Here we present observations for a simple experiment: are Be stars more likely
than normal B stars to be in a binary system? To this end, we obtained Adaptive
Optics (AO) data of a matched sample of 40 B and Be stars using an 8 meter class
telescope.

\begin{figure}
\begin{center}
\includegraphics[width=0.45\textwidth]{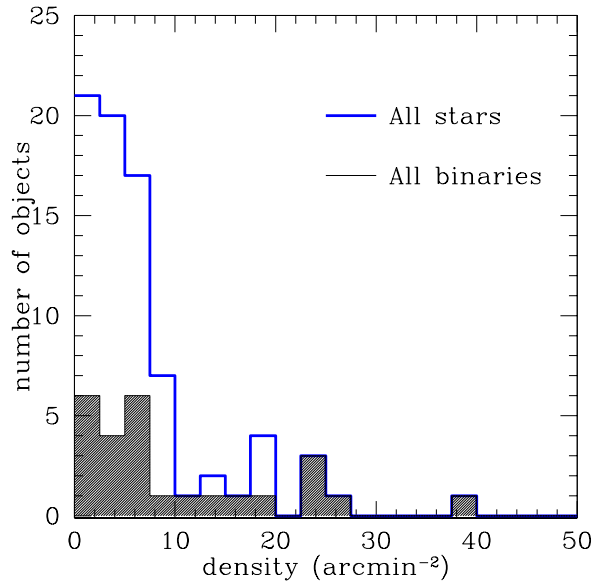}
\caption{The number of 2MASS sources per square arcminute around the
target objects. A density of 20 arcmin$^{-2}$ corresponds to an average
of one background object per field of view of 13.6$\times$13.6
arcsec$^2$. All targets in fields with source densities exceeding this value have
indeed more than one object in the FoV and were initially flagged as
having a binary companion, but are not included in the final sample. 
\label{densfig}}
\end{center}
\end{figure}

\section{Observations and Data Reduction}

\subsection{Target selection}

In order to arrive at an unbiased, bright sample of Be stars, we used the
Bright Star Catalogue \citep[BSC]{bsc} as a starting point. We selected on
spectral type and luminosity class (BIVe and BVe), and on Declinations that
can be covered by the ESO-VLT.  There are 71 BVe and BIVe stars in the BSC and
its Supplement located below +10$^{\rm o}$ Declination.
As comparison, we selected all BV and BIV stars with Declination smaller than
+10$^{\rm o}$. This returned 640 objects, consistent with the notion mentioned
above that about 15\% of all B stars are of Be spectral type. A random
sub-sample of B stars was selected from this master sample to match the number
of Be stars to be observed.  As Be stars predominately occur for earlier B
spectral types, a slight preference was given to the earlier type B stars in
this selection.  The typical {\it K} band magnitudes of the Be stars are
around 5-7.  The intrinsic {\it K} magnitudes for M0 dwarfs are (only) 6-9
magnitudes fainter than those of B9V and B0V stars respectively.  Thus, for a
limiting {\it K-}band magnitude of 12.5 almost the entire main sequence down
to the M dwarfs can be probed.  A total of 40 B and 39 Be stars was observed
using adaptive optics (AO) on Yepun, one of UTs of the Very Large Telescope
(VLT) at the European Southern Observatory (ESO) in Chile.  No selection on
binarity or brightness was applied, instead, the targets were chosen for their
observability in terms of their Right Ascension, their location on the
sky. The targets are presented in Table~\ref{logtarget}.

\subsection{Observations}

The observations were carried out in service mode during 9 nights
between 24 March 2005 and 5 June 2005. The observations were conducted
with the AO system NACO \citep{lenzen_2003,rousset_2003}.  In order to
achieve the best resolution images, data were taken in the {\it K}
band, where the diffraction limit is 0.057 arcsec or 57 milli-arcsec
(mas).  NACO has a 1024$\times$1024 pixel array, with a pixel size of
27 $\mu$m. The finest plate scale of 13.27 mas$/$pix was used, giving a
field of view of 13.6 arcsec $\times$ 13.6 arcsec.

\begin{figure*}
\begin{center}
\includegraphics[width=0.95\textwidth]{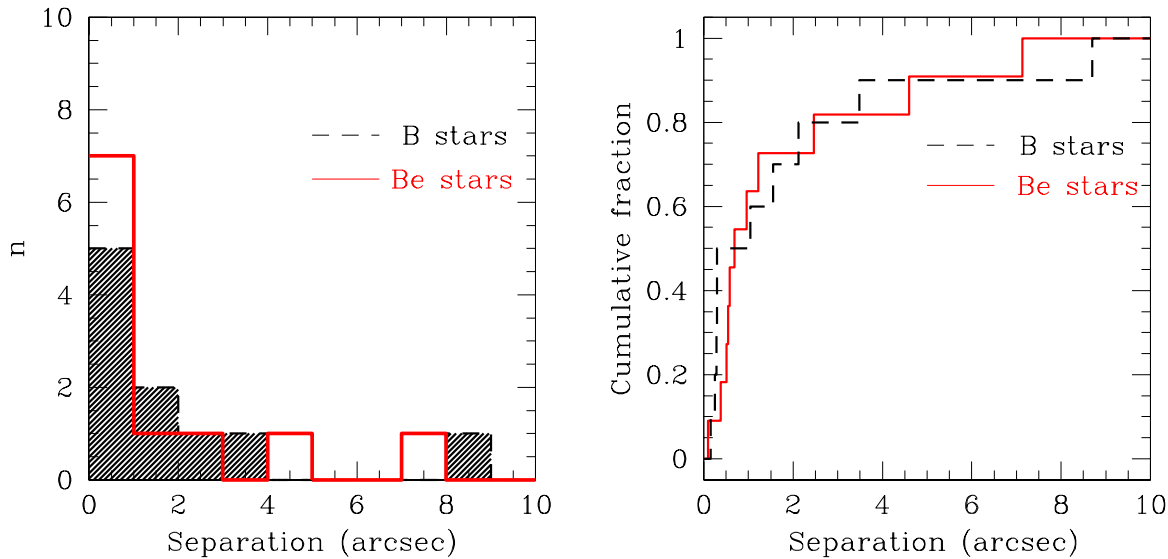}
\caption{The binary separations (left) and their cumulative distributions
  (right) of the B and Be binaries respectively.  The Be star data are shown
  with red solid lines, those of the B stars as black dashed lines. Following
  KS statistics, it can be concluded that the samples are drawn from the same
  parent distribution (see text).
 \label{sepfig}}
\end{center}
\end{figure*}

As the targets are very bright, the NB2.17 narrow band filter was
used. This filter is centered on Br$\gamma$ with a central wavelength
of 2.166 $\mu$m and a full-width at half maximum (FWHM) of 0.023 $\mu$m.
The contribution of Br$\gamma$ emission to the total flux from the Be
stars is negligible - a typical emission line with an Equivalent Width
of 2-3 $\rm \AA$ \citep[see e.g.][]{persson_1985} would contribute
less than $\approx$ 1\% to the total flux received by the filter.  For
each target star, 5 frames were taken in the auto-jitter mode with a
box size of 5 $\times$ 5 arcsec. The integration time was dependent on
the brightness of the target, and was in most cases 12 seconds per
jitter position. For the very brightest sources, a Neutral Density
filter was employed.  The seeing was always better than 1.4
arcsec, and airmasses were below 1.2, in order to achieve a good AO
correction. The targets were used as AO reference stars.

The data were reduced with the ESO pipeline reduction software.  Dark frames
were produced on the same night as each observing run, with the same
integration time as the science frames produced.  Dark subtraction and
flat field division were performed on individual frames by the
pipeline, before the 5 calibrated frames were combined.  This was
performed by automated detection of the target, determining an offset
then registering and stacking of the 5 calibrated frames.  All the raw
data were checked by eye to investigate whether the automated
detection did not produce artificial binaries, and in the few
ambiguous cases, the shift-and-add procedure was done by hand.

Typical FWHM of 0.09 arcsec were measured on the final images, and the point
spread function's (PSF) widths ranged between nightly averages of 0.07 arcsec
and 0.10 arcsec.  The typical Strehl ratio achieved was 30\%, with a fair
fraction of targets having Strehl ratios exceeding 45\%.

\subsection{Binary Detection}

The raw and reduced images were visually inspected to identify binary
companions.  The binary separation and position angle were determined
by fitting a two-dimensional Gaussian function to both spatial
profiles.  The maximum separation found was 8 arcsec, which is about
the maximum possible value in our square field-of-view with the target
source positioned in the centre of the frame and a 5 arcsec jitter
offset. The minimum separation was 0.1 arcsec, at the resolution limit
of our data.  The difference in {\it K}-band magnitude (or rather,
difference in magnitude of the light through the NB filter, $\Delta
m$, which is very close to $\Delta K$), was measured for each
component using either aperture photometry, or by estimating the total
counts of each star, based on a two-dimensional Gauss fit of the
spatial profiles.  In the case of more than one companion, the
magnitude difference was found for the closest of the companions. In
two cases it proved difficult to obtain a reliable estimate of the
secondary's magnitude and the entry is left blank in
Table~\ref{restarget}. The largest, reliable, magnitude difference was
7.3 magnitude.  As a check on the quality of our data, the parameters
of known binaries were compared to our results.  For example, HR 3745
was identified as having a separation of 0.55 arcsec and a PA of
$275^\circ$ by \cite{hartkopf_1996}. Here we find a separation of 0.55
arcsec. and a PA of $281^\circ$. The difference in position angle is
not necessarily an indication of the errors involved as it can well be
due to orbital motion.

We find 14 visual binaries among the 40 observed B stars and 13 visual
binaries among the 39 Be  stars observed. Taken at face value, the
binary fractions are therefore 35 $\pm$ 8 \% (70\% confidence interval) for
the B stars and 33 $\pm$ 8 \% for Be stars.

The numbers for B and Be stars are very close and comfortably within the
uncertainties.  However, it is not obvious that these concern physical
binaries or visual binaries whose secondaries have no physical association
with the target stars. The best way to test this is to obtain second epoch
data to investigate whether the secondary stars share a common proper motion
or not. In the absence of such data, we consider the following as a proxy for
the possibility of finding a chance binary companion.

The chances of finding one star in the field of view are large when the
stellar density is at, or exceeds, a value of 1 star per field of view of 13.6
$\times$ 13.6 arcsec$^{2}$ corresponding to 19.5 stars per arcmin$^{2}$. A
measure of the local surface density of stars was obtained from the 2MASS
Point Source Catalogue \citep{2mass} by tallying up all objects within a
circle with a radius of 1 arcmin from the target objects. The frequency of the
stellar densities towards the sample is shown in Fig.~\ref{densfig}. Most
objects lie in regions with stellar densities less than 10 per arcmin${^2}$,
less than half a star per field of view, and the presence of a chance
companion is not expected. However, all six objects which lie in densely
populated areas on the sky (densities larger than 20 arcmin$^{-2}$), are found
to have a companion. It is much more likely that these objects are normal
field stars that happen to be in the field of view rather than physical binary
stars, and we will proceed without those objects.

\section{Results}

\subsection{Binary fraction}

After selecting on the objects with a local field star surface density
less than 20 arcmin$^{-2}$, we remain with 11 Be binaries out of 37
targets (30 $\pm$ 8 \%) and 10 B binaries out of 36 targets (29 $\pm$
8\%). The B and Be binary fractions are remarkably similar.  In the
remainder of this study, we will continue the analysis with these 21
binary systems. For completeness we note that if the density criterion
were lowered to 12.5 or 15 to exclude the next two objects, the
statistics remain the same, but now both samples would be reduced by
exactly one object.  HR 5661 (a Be star with a local source density of 16
arcmin$^{-2}$) and HR 6316 (a B star with a local source density of 18
arcmin$^{-2}$) remain in the sample.

In the following, we investigate and compare basic properties of the binary
systems such as separations and flux differences - which directly translate
into mass ratios - between the primaries and secondaries.  The binary systems
and their parameters are listed in Table~\ref{restarget}.


\begin{figure}
\begin{center}
\includegraphics[width=0.45\textwidth]{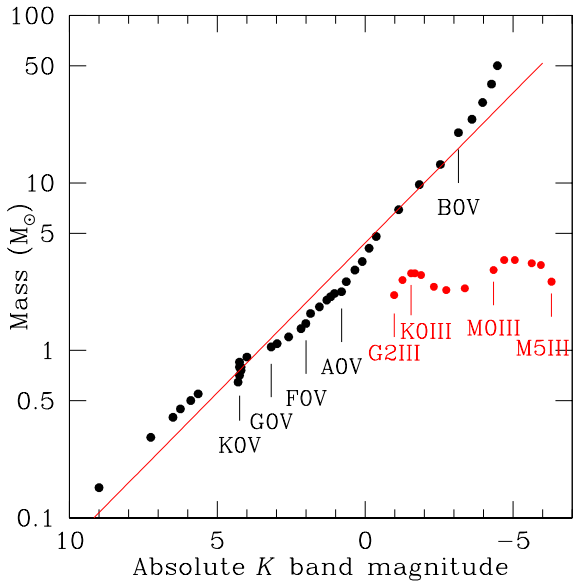}
\caption{The masses of Main Sequence stars (black dots) and giant
stars (red dots) plotted against their intrinsic {\it K} band
magnitudes. There is a one-to-one relation between these parameters for
Main Sequence stars, allowing the mass ratio of a binary to be derived
based on the {\it K} band magnitude difference only. As we know the
spectral type of the primary, we can in most cases rule out the secondary to be a
giant or supergiant, as these would be brighter than the primary,
which is not observed.
\label{sptpfig}}
\end{center}
\end{figure}

\begin{figure*}
\begin{center}
\includegraphics[width=0.95\textwidth]{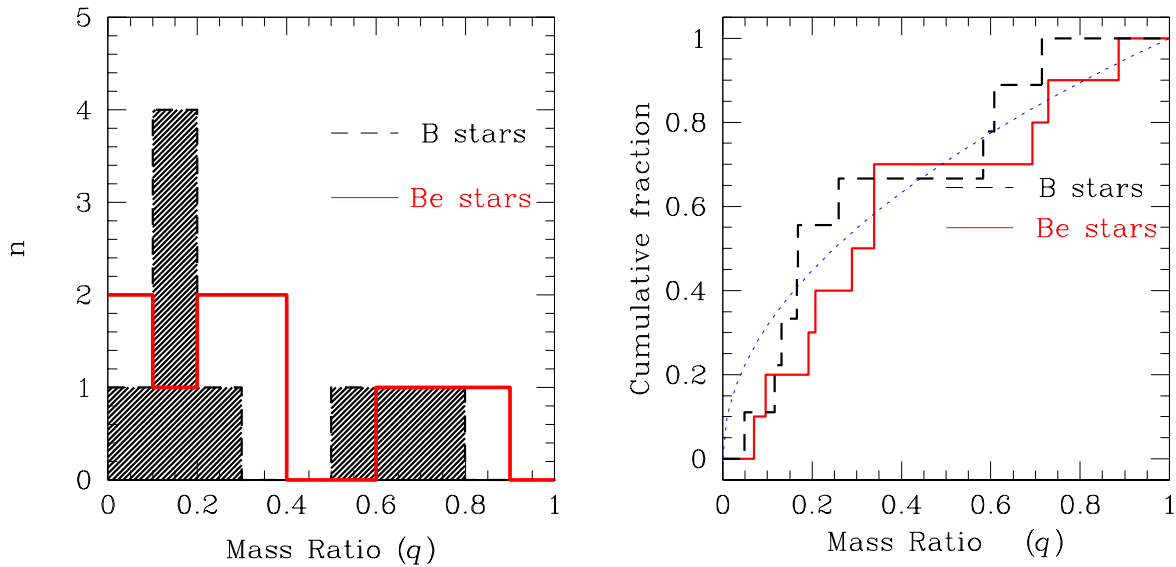}
\caption{
As Fig.~\ref{sepfig}, but now for the mass ratios of the binary
systems. The mass ratios of the B and Be star binaries are not
markedly different and according to KS statistics drawn from the same
parent population. The dotted line is not a fit to the data, but the
distribution found by \citet{shatsky_2002} for a larger sample of B
stars. 
\label{mratiofig}}
\end{center}
\end{figure*}

\subsection{Binary separation}

Here we compare the binary separations for all B and Be stars. A histogram of
the separations is shown in Fig.~\ref{sepfig}. The Be-star separations are
slightly more centrally peaked than the B star separations. Although it is not
a very significant result, we would wish to draw attention to the fact that
this may be expected. The Be stars are rarer than B stars and they are thus,
on average, further away. This is corroborated by the parallaxes of the sample
objects.  Using the Hipparcos parallaxes taken either from \citet{hipparcos}
or \citet{vanleeuwen_2007}, we find that the average parallax of the
B-binaries is 4.9 milli-arcsec (with a generous spread of 2.7 mas) while the
average parallax of the Be stars is slightly
lower at 4.4 mas (with a $\sigma$ of 2.8 mas). Should the intrinsic separation
distribution be similar, then it is to be expected that the Be stars'
separations (in arcsec) are smaller, or somewhat more centrally peaked than
for the B stars. We note however, that this is not a large effect.  We also
performed a 2-sample Kolmogorov-Smirnov (KS) test. The cumulative distribution
is shown in the right hand panel of Fig.~\ref{sepfig}.
The KS statistic indicates that the hypothesis that the Be and B binary
separations are drawn from different separation distributions can be rejected
at the 74\% level. This is not overly conclusive, but the sample is large
enough to have identified large differences in the distribution of
separations, if there would have been any.

As conclusion of this section, we find that the Be binaries are
similar to the B binaries in terms of separation.

\subsection{Mass ratios}

Here we consider the masses of the components in the binary
systems. In principle, we can infer the masses of the primaries from
their listed spectral types using published calibrations, and derive
spectral types of the secondaries using the flux differences and by
comparing with the intrinsic magnitudes listed for the different
spectral types, we can then determine the secondaries'
masses. Although this is a standard method, it is fairly elaborate and
introduces various errors. These comprise the error in spectral
typing, the mass and magnitudes range for a given spectral type
(cf. e.g. the width of the Main Sequence, \citet{hipparcos}, see also
\citet{oudmaijer_1999} for an in-depth study of K0V stars) and affect
the final result.

These issues are mostly avoided in the present study by simply using the {\it
  K} band magnitude difference. This provides an excellent opportunity to
derive the mass ratio with no dependence at all on the spectral type of either
component.  It turns out that there is a strong relation between the intrinsic
{\it K} band magnitude and the mass of a main sequence star.  This is shown in
Fig.~\ref{sptpfig}, where the mass of a star (taken from \citealt{str_kur}) is
plotted against the intrinsic {\it K} band magnitude which was computed using
the intrinsic {\it V} band magnitudes also taken from \cite{str_kur} and the
intrinsic {\it V $-$ K} colours due to \citet{koornneef_1983}.  The relation
for Main Sequence stars is close to a straight line which can be represented
as

$$ {\rm log}(M) = -0.18K + 0.64 $$

\noindent
with $M$ in solar units, and a formal uncertainty on the slope of
0.005. This in turn translates into a mass ratio $q$ - without the need
of knowing the masses separately:

$$ q = 10^{-0.18\Delta K} $$

\noindent
with $\Delta K$ the magnitude difference between both objects in our data.

By virtue of the rather linear relationship between {\it K} band flux and
stellar mass, the mass ratio can be determined independent of prior knowledge
of the stellar spectral types and luminosities while it is independent of
reddening too. The only assumption we tacitly make is that both the primary
and secondary stars under consideration are on the Main Sequence. This may
seem obvious, but we note that many known binaries with a hot O, B type dwarf
star have a cool, G, K or M-type giant secondary (see
e.g. \citealt{ginestet_2002}). In our sample, HR 2577, an unresolved
spectroscopic binary with a K2II companion listed in Table~\ref{logtarget} is
a prime example. In most such binary stars, the B-type component dominates at
optical wavelengths, while the secondary dominates at longer wavelengths,
simply because cooler stars are redder. In contrast, the current study is
conducted in the infrared already and the brightest binary component is the
previously known B-star whose catalogued {\it K} band magnitide is consistent
with the optical brightness, while the secondary is fainter.

For reference, we have also plotted the mass of a selected number of giant
stars against their intrinsic {\it K} band magnitudes in Fig.~\ref{sptpfig}.
The cool G, K and M-giants are brighter than most B stars up to a spectral
type of B3V.  This means that for all stars with spectral types of B3 or later
in our sample, we can be fairly certain that the companion is not a giant as
the primary star, the B star, is the brightest. Indeed, only for B0V-B3V stars
there would be a chance that the (fainter) companion is a giant rather than a
main sequence dwarf star. In such a case, the magnitude difference will be at
most 2 magnitudes. Inspection of Table~\ref{restarget} however reveals that
almost all binaries with a B0V-B3V primary have a magnitude difference larger
than 2. We conclude that it is very likely that the secondary stars are main
sequence stars and that we are therefore justified in applying our {\it
  K}-mass relation to determine the mass ratio of the systems.

One correction needs to be made for the Be stars, the free-free emission from
the ionized disk increases with wavelength, and can give a slight excess
emission at the {\it K}-band (e.g. \citealt{dougherty_1991}). To arrive at the
intrinsic magnitude of the star itself, we need to subtract the disk
contribution from the {\it K} band flux. This is done in the standard manner
by dereddening the {\it V} band magnitude (using intrinsic colours for the
respective spectral types from Schmidt-Kaler in \citet[Chap. 4]{Landolt}) and
computing the predicted {\it K} band magnitude using the {\it V $-$ K} colours
as outlined above. The difference between the predicted and observed {\it
  K}-band magnitude is the excess. In two cases, HR 6712 and HR 5193, the
excesses were negative by up to half a magnitude and thus unphysical. These
stars are variable and using the optical colours from SIMBAD rather than from
the BSC solved this situation. The ``excesses'' computed for the normal B
stars averaged 0.01 magnitude with a scatter of 0.1 magnitude, which gives an
indication of the uncertainties involved in this exercise. As all magnitude
differences between the companions are much larger than 0.3 magnitudes, this
error does not affect the resulting mass ratio. In fact, only two Be stars
have computed excesses which are much larger than 0.3 magnitude, these are HR
2142 (0.6 mag) and HR 2356 (0.8 mag). After correcting for the excess
emission, we can then compute the photospheric magnitude difference and the
mass ratio $q$.

In Fig.~\ref{mratiofig} the results for the mass ratios are presented in a
similar fashion as for the separations in the previous figure. Note that the
two objects for whose close companions we were unable to retrieve accurate
photometry are not plotted. The left hand plot shows the distribution of mass
ratios, which range from 0.07 to 0.9. Based on the relatively low numbers, it
is hard to describe this distribution, but it would seem that the
distributions are similar and rather flat for the B and Be stars. It does not
rise steeply to low masses as would be the case if the secondary masses were
randomly sampled from the Initial Mass Function (see
e.g. \citealt{Wheelwright_2010}).  The cumulative plot in the right hand panel
confirms that there is not a significant difference between the two
distributions. The KS test returns a probability that the samples are drawn
from different populations can be rejected at 50\%.  As with the separations,
if the mass ratios would have been very different, the current sample is large
enough to highlight this. It therefore seems likely that the mass ratios of
both the B and Be type binaries too are sampled from the same parent
distribution.  For reference the function $f(q) \propto q^{-0.5}$ is
overplotted as a dashed line. This is not a fit to our data, instead it is
what \citet{shatsky_2002} found to best represent the mass ratio they
determined for a large sample of B-type binaries. Our data are consistent with
their data, which sample a similar separation and magnitude range.

In summary of this exercise, there is no difference  between B and
Be binaries as far as their companion masses is concerned.

\begin{table}
\begin{center}
\begin{tabular}{lllrlll}    
\hline
\hline
  HR  &  SpTp  & $\rho$  & $\Theta$ & $\Delta \, m$  &  {\it K} & $q$ \\
      &        & (arcsec) & {$^o$} & (mag) & (mag)   & \\
\hline
 1753 &   B3V &   0.28 &    -65 &    4.3 &   6.79  & 0.17\\ 
 1892 &   B1V &   0.15 &    -32 &    1.2 &   5.06  & 0.61\\ 
 2056 &   B5V &   1.04 &    198 &    5.2 &   5.25  & 0.12\\ 
 3990 &  B3IV &   0.29 &    -56 &       &   5.10  &     \\ 
 4940 &   B5V &   1.55 &    267 &    3.2 &   5.05  & 0.26\\ 
 5559 &   B9V &   2.12 &    278 &    0.8 &   5.68  & 0.72\\ 
 5708 & B2IV-V &   0.29 &    228 &    1.3 &   4.13 & 0.58\\ 
 6316 &   B8V &   8.70 &     27 &    7.3 &   5.26  & 0.05\\ 
 6960 & B2IV-V &   0.25 &    178 &    4.3 &   5.54 & 0.17\\ 
 7709 &   B1V &   3.49 &    150 &    4.9 &   6.95  & 0.13\\ 
\hline
 2142 & B2Vne &   0.58 &    269 &    4.6 &   4.78  & 0.19\\ 
 2231 &  B6Ve &   0.39 &    -83 &    3.8 &   6.38  & 0.21\\ 
 2356 &  B3Ve &   7.14 &    133 &    1.7 &   4.08  & 0.70\\ 
 2412 &  B9Ve &   2.46 &    221 &    2.7 &   5.72  & 0.34 \\ 
 2690 & B2IVe &   0.69 &    228 &    6.0 &   5.64  & 0.10\\ 
 3745 &  B6Ve &   0.55 &    281 &    1.0 &   6.09  & 0.73\\ 
 3946 &  B4Ve &   0.51 &     -9 &    3.2 &   6.23  & 0.29\\ 
 5193 & B2Vnpe &   4.60 &    -57 &    6.4 &   4.01 & 0.07\\ 
 5661 & B0.5Ve &   1.22 &    164 &    2.6 &   5.94 & 0.34\\ 
 5683 &  B8Ve &   0.96 &    -55 &    0.3 &   4.43  & 0.89\\ 
 6712 &  B2Ve &   0.10 &    154 &        &   5.03  &     \\ 
\hline
 \hline
\end{tabular}
\caption{The final binary detections, with objects in dense fields
  excluded (see text). Listed are the separations and position angles (measured East of
  North), as well as the magnitude differences. Typical errors are of order
  0.05 arcsec, 1$^o$ and 0.1 magnitude respectively. {\it K} band magnitudes
  are taken from 2MASS.  The mass ratios $q$ are derived in the text.
\label{restarget}
}
\end{center}                                                  
\end{table}

\section{Discussion and final remarks}

We have observed a total of 79 B and Be stars at high spatial resolution to
study their binary properties. This is the first systematic comparative and
deepest imaging study into the binarity of Be stars published. The
observations probe scales down to less than 0.1 arcsecond and allow companions up
to 10 magnitudes fainter to be detected. In the process, we derived a simple
relation between the {\it K}-band magnitude difference and the mass-ratio of
the binary components under the assumption that both components are on the
Main Sequence. The latter is justified in our sample.  After excluding a few
visual binaries that are located in regions of high stellar density, we find
that :

\begin{enumerate}

\item There are 10 detected binaries out of 36 B stars and 11 detected
binaries out of 37 Be stars (corresponding to binary fractions of 29
$\pm$ 8 \% and 30 $\pm$ 8\% respectively). The binary fractions are
nearly identical.

\item The separations of the B and Be binary systems follow the same
cumulative distributions.

\item 

The mass ratios for the 9 B and 10 Be binaries for which we have
differential photometry have similar distributions.  Like for the
separations, the samples appear to be drawn from the same parent
distribution.

\end{enumerate}

\noindent
This leads to our main conclusion :

\begin{itemize}

\item The binary fraction, and the binaries' properties, of Be stars
in the separation range 20 - 1000 AU are very similar to those of a
matched sample of B stars. This is a statistically robust result and
as both samples were observed with the same observational equipment,
and therefore were subject to the same observational biases, we can
rule out binary companions at separations of 20 AU or more as being
responsible for the Be phenomenon.

\end{itemize}

Returning to the original motivation of the survey, the previously found large
fraction of wide Be binaries has been reduced by a factor of 2 due to improved number
statistics.  More importantly, the comparison with normal B stars now allows
us to firmly conclude that there is no need, based on this study, to explore
new models involving wide binaries for the Be phenomenon. 

Closer binaries than 20 AU can only be found with methods that have
even higher spatial resolution than here (diffraction limited imaging
at an 8 meter telescope in the {\it K} band), or with radial velocity
studies. A systematic imaging study at even higher resolution (indeed
at any resolution) such as this one has not been performed, but the
systematic spectroscopic study by \citep{abt_1978} does not reveal a
large percentage of close binaries.  Indeed, on the contrary, they
find a lower close binary fraction among Be stars than B stars.
Therefore, binarity can not be responsible for all Be stars.

Finally, a word on the similar mass ratios and binary separations. The mass
ratio and separation of a binary system are the result of the initial
conditions governing the formation of both components and their subsequent
evolution. The stars in our sample are field stars, which means that if they
were formed in clusters, as most massive stars do, these clusters have
dispersed long since.  Given that the separations of the B and Be binaries
under consideration are fairly wide, the objects were most likely formed in
low density clusters or even in isolation. This is because large separation
binaries will be destroyed early in the evolution of dense clusters, while
close, more strongly bound binaries are most likely to survive throughout the
entire evolution \citep{kouwenhoven_2009,parker_2009}.  It is beyond the scope
of this paper to derive the precise cluster conditions, but it would appear
that both the evolutionary history and the birthplaces of the B and Be stars
are very much the same.  It can be concluded from our data, therefore, that the
Be phenomenon does not result from a substantially different star formation
mechanism or conditions compared to those for normal B stars.

\section*{Acknowledgments}

RDO is grateful for the support from the Leverhulme Trust for awarding
a Research Fellowship.  AMP acknowledges support from The Rothschild
Community of Excellence Programme. We thank the referee, Helmut Abt,
for his perceptive comments. It is a pleasure to thank Ben Davies,
Hugh Wheelwright and Willem-Jan de Wit for their comments on an
earlier version of the manuscript. We thank Simon Goodwin for
insightful discussions and Simon Baber for his expert help in the
final stages of this project.

\bibliographystyle{mn2e}
\bibliography{bebin}

\end{document}